\documentclass[showpacs,prb,manuscript,aps]{revtex4} \draft
\usepackage[pdftex]{graphicx}
\usepackage[tight,footnotesize]{subfigure}

\begin{document}

\title{ Semiconductor to Metal Transition, Dynamical Stability and Superconductivity of Strained
Phosphorene }

\author{G. Q. Huang$^{1,2}$, Z. W.  Xing$^{2,3}$}
\affiliation{$^{1}$Department of Physics, Nanjing Normal University,
Nanjing 210023, China\\
$^{2}$National Laboratory of Solid State Microstructures, Nanjing
University, Nanjing 210093, China\\
 $^{3}$Department of Materials Science and Engineering,  Nanjing University, Nanjing 210093, China}

\begin{abstract}
{Very recently, field-effect transistors based on few-layer phosphorene  crystals with thickness down to a few
nanometres have been
 successfully fabricated, triggering  interest in this new functional
two-dimensional material. In this work, we apply the first-principles
calculations to studying the evolution of electronic
structures and lattice dynamics  with vertical strain for monolayer and bilayer
phosphorenes. It is found that,   by changing the thickness of
phosphorene or the strain applied on it, its band gap width can be well tuned, and there will appear a
transition from semiconductor to metal.  In particular, the bilayer phosphorene may become a good BCS superconductor
by adjusting the interlayer distance, in which the interlayer
van der Waals coupling is favorable to  the dynamical stability against strain.   }

\end{abstract}

\maketitle

\section{Introduction}
Black phosphorus (BP) is the most stable allotrope of the element
phosphorus and has many interesting physical properties. BP can
server as the electrode material for Lithium-Ion
batteries.\cite{1,2} Several structural phase transitions were found
under pressure accompanied by semiconductor-semimetal-metal
transition.\cite{3,4} Furthermore, it was  reported that the BP
single crystal shows superconductivity with $T_{C}$ higher than 10K
under high pressure.\cite{5,6} The structure of BP consists of
puckered double layers, which  are held together by weak van der
Waals (vdW) force. This peculiar layer structure permits to employ
mechanical exfoliation to extract thin black phosphorus from a bulk
crystal. Recently, the few-layer black phosphorus (phosphorene) had
been successfully exfoliated and used them to create field-effect
transistors, which exhibit large current on-off ratios and high
mobilities.\cite{7,8,9} The successful fabrication  of this novel
two-dimensional (2D) semiconducting material was soon paid amounts
of attention and predicted to have a great potential for practical
applications.\cite{10,11}

\par
Bulk BP is a direct-gap semiconductor with a 0.33 eV band
gap.\cite{12} However, the band gap of  phosphorene is predicted to
be highly sensitive to its thickness and strain. Previous
first-principles calculations show that the band gap ranges from 1.5
eV for a monolayer to 0.6 eV for a 5-layer.\cite{13} A 3$\%$
in-plane strain can change phosphorene from a direct-gap to an
indirect-gap semiconductor,\cite{8} while a vertical compression can
induce the semiconductor to metal transition.\cite{14} Strain is an
effective method to tailor electronic properties for 2D
semiconductors. For monolayer phosphorene (MLP) and bilayer
phosphorene (BLP), whether there are different strain effects on
electronic properties is one of the motivation of this study.
Furthermore, due to the interlayer coupling bonded by weak van der
Waals force,  it is also interesting to see how bilayer phosphorene
responses to the change of interlayer distance.

\par
While there have been substantial works on the electronic properties
of phosphorene, investigations on the vibrational properties of
phosphorene are scarce. The Raman spectra was measured for the BP
flake with thicknesses ranging from 1.6 nm to 9 nm.\cite{15} The
phonon dispersion of MLP was calculated by first-principles
calculations.\cite{10,16} The lattice vibrational modes of MLP with
in-plane strain were also reported by Fei \textsl{et al.} through
first-principles simulations.\cite{16} However, phosphorene has an
anisotropic structure, whether it has different dynamical behavior
for phosphorene with out-of-plane strain? This problem is our
concern. We will  pay particular attention to the dynamical
stability of strained phosphorene   in order to better use them for
practical application.

\par
In this work, geometric structure, electronic structure and lattice
dynamics of MLP and BLP  under vertical strain are studied and
compared by first-principles calculations. The calculated results
show that the evolution behavior of band structure for MLP and BLP
under vertical strain is different. The dynamical stable range of
the strained BLP is wider than that of the strained MLP.
Furthermore, we also find that BLP may become a good BCS
superconductor by adjusting the interlayer distance.

\section{ Computational Details}
\par
The first-principles calculations have been performed within the
density functional  theory through the PWSCF program of the
Quantum-ESPRESSO distribution.\cite{17} The ultrasoft
pseudo-potential and general gradient approximation (GGA-PBE) for
the exchange and correlation energy functional are used with a
cutoff of 30 Ry for the expansion of the electronic wave function in
plane waves. MLP and BLP are modeled using the slabs, which are
separated by vacuum layer with the thickness of about 35 $\AA$. For
the electronic structure calculations, the Brillouin zone
integrations are performed with a (8,10,1) wavevector $k$-space grid
by using the first-order Hermite-Gaussian smearing technique. Within
the framework of the linear response theory, the dynamical matrixes
are calculated for special $\bf{q}$ points in the two dimensional
irreducible Brillouin zone and are Fourier interpolated throughout
the full Brillouin zone.  The dense (24,30,1) grid are used in the
Brillouin zone integrations in order to produce the accurate
electron-phonon (EP) interaction matrices. As BP presents a laminar
crystal structure, vdW  correction proposed by Grimme (DFT-D2) is
included in our calculations.\cite{18}

\section{ Results and discussion}
\subsection{ Geometric structure under vertical strain}

BP has a layered orthorhombic structure. Each sheet consists of an
atomic double layers known as a puckered layer.  The stacking of the
puckered layer for BP is \textsl{ABAB} $\ldots$ sequence with weak
van der Waals bonding between layers. The edge of the puckered
hexagon of A layer is located in the center of the puckered hexagon
of B layer and vice versa. The top and side views of the atomic
structure of a puckered layer are presented in Figs. 1 (a) and (b),
respectively. Each layer may be viewed as consisting of armchair and
zigzag chains along $x$ and $y$, respectively. Each phosphorus atom
is covalently bonded to three neighbors  within the puckered layer.
Our optimized bond lengthes, bond angles (as labeled in Fig. 1) and
lattice constants for BP are listed in Table 1. For comparison, the
results without vdW correction for  bulk BP are also listed in Table
1. It can be clearly seen that the inclusion of vdW interaction
changes $a$ and $b$ parameters slightly, but reduces the c parameter
substantially (about 0.6${\AA}$). The overall best agreement with
experimental\cite{19} and previous computational results\cite{20} is
achieved with inclusion of vdW correction.

\begin{figure}
\vskip 1in
\centering
\includegraphics[width=4.in]{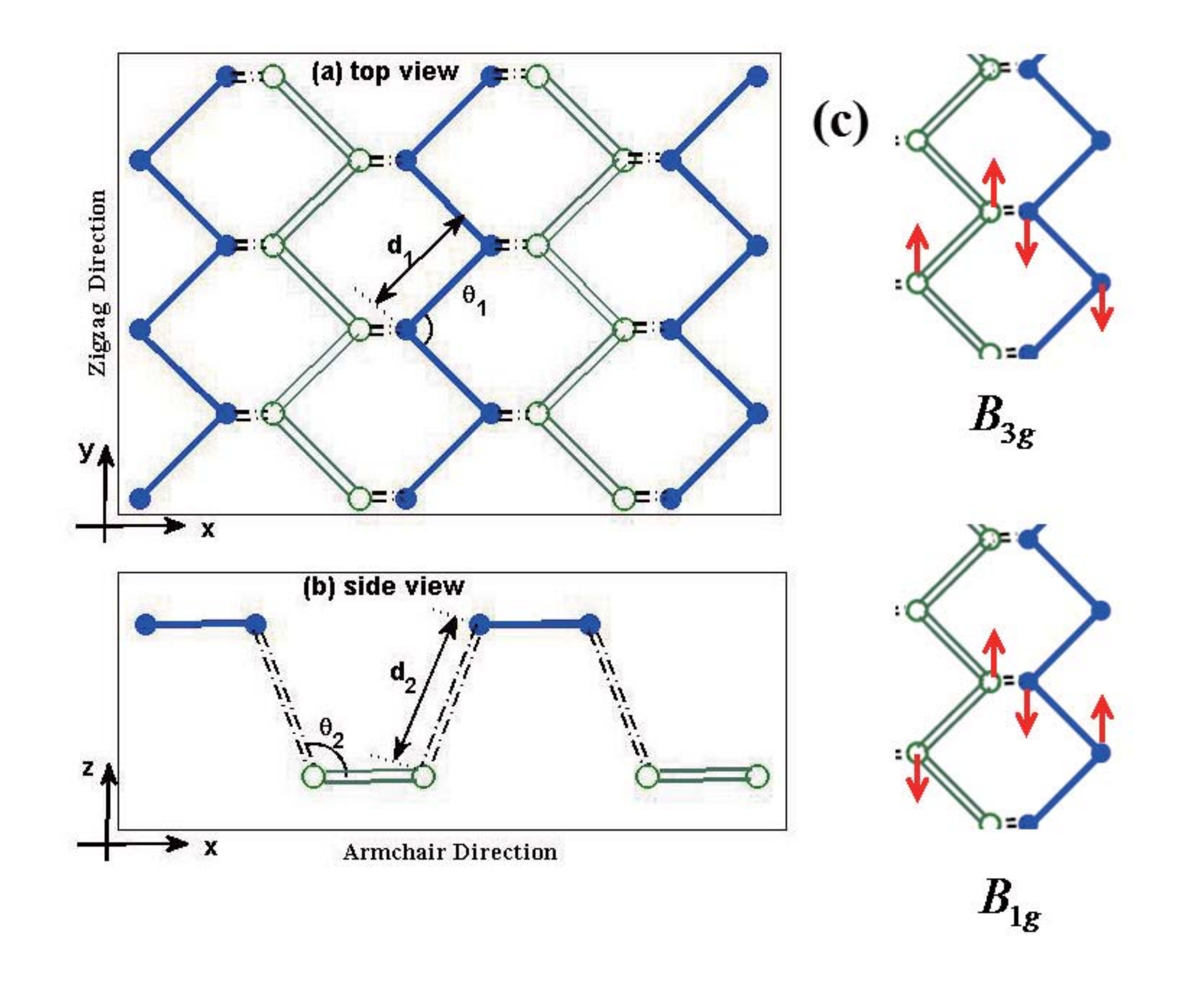}
\caption{The top (a) and  side views (b) of the atomic structure of
 phosphorene.  (c) Vibrational pattern of $B_{3g}$ and
$B_{1g}$ modes.}
\label{fig.1}
\end{figure}

\par
For the free MLP and BLP (see Table 2), lattice constant $b$ and
bond lengthes are very close to those in the bulk BP, while  $a$ and
bond angles $\theta_{2}$ are larger than those for the bulk BP. This
fact shows that the increase of lattice constants $a$ is not due to
the change of bond lengthes, but is due to the flattening of
puckered layer. Next we concentrate mainly on the strained
phosphorene. The vertical strain is modeled by constraint $z$ for
the outmost layer atoms, the unit cell and other atomic positions
are then relaxed. The strain is defined as
$\sigma=\frac{h-h_{0}}{h_{0}}$, where $h$ and $h_{0}$ are the
thickness of the strained and the free phosphorene, respectively.
The positive (negative) of $\sigma$ corresponds to the tensile
(compressive)   strain. The obtained geometric parameters for MLP
and BLP under vertical strain are listed in Table 2.

\par
For the  MLP, when tensile (compressive) strain is applied in the
$z$ direction, the lattice constant $a$ decreases (increases)
substantially. While the lattice constant $b$ and the bond length
$d_{1}$ show only a weak variation under strain,  reflecting the
rigidity of the strong covalent bonding along the zigzag direction.
So the anisotropic geometric structure along armchair ($x$ ) and
zigzag ($y$ ) directions result in their different responses to the
vertical strain. With the decrease of thickness ($h$) of
phosphorene, the bond length $d_{2}$ decreases and the bond angle
$\theta_{2}$ increases, reflecting the flattening of the puckered
layer. It is need to mention that  the range of strain studied here
is not too large, the puckered character of  structure still
remains. The previous calculations by Rodin \textsl{et al.}\cite{14}
showed that the monolayer approaches a plane square lattice
configuration under severe compression.

\par
For the  BLP, when tensile strain is applied, the distance between
puckered layers increases, while the covalent bonds within the
puckered layer change slightly. And we also find that the geometric
parameters of BLP are tend to those in the MLP when tensile strain
is increased to $40\%$. This result is reasonable, since the
coupling between puckered layers may be ignored at so large
interlayer distance (about 6 ${\AA}$). With increase of  compressive
strain applied, the change of  geometric parameters of BLP is
qualitatively same as  in the case  of MLP but with smaller
magnitude. Unlike to that for the MLP, the lattice constant $a$ of
the BLP always increases whatever compressive or tensile strain is
applied along z direction. This peculiar property is same to that
observed in the bulk BP.\cite{21} This result implies that BLP has
the same unusual mechanical response as bulk BP.

\subsection{ Electronic structures under vertical strain}
The calculated band structures of  the strained MLP and BLP are
shown in Fig. 2. For the free MLP (Fig. 2c), the band dispersion
along the major high symmetry directions compares very well with the
previous calculations by Rodin \textsl{et al.}.\cite{14} It is a
direct band gap semiconductor with the band gap energy about 0.83
eV. The top of the valence band and the bottom of the conduction
band have predominantly $p_{z}$ character mixed with $p_{x}$ and $s$
character. The band structure of  the free BLP (Fig. 2h) has similar
shape to that of MLP but with smaller gap energy (0.40 eV). The
photoluminescence measurements\cite{15} and previous
calculations\cite{13} indicated a thickness dependent band structure
of the phosphorene.

\begin{figure}[!t]
\centering \subfigure {
\includegraphics[width=2.in]{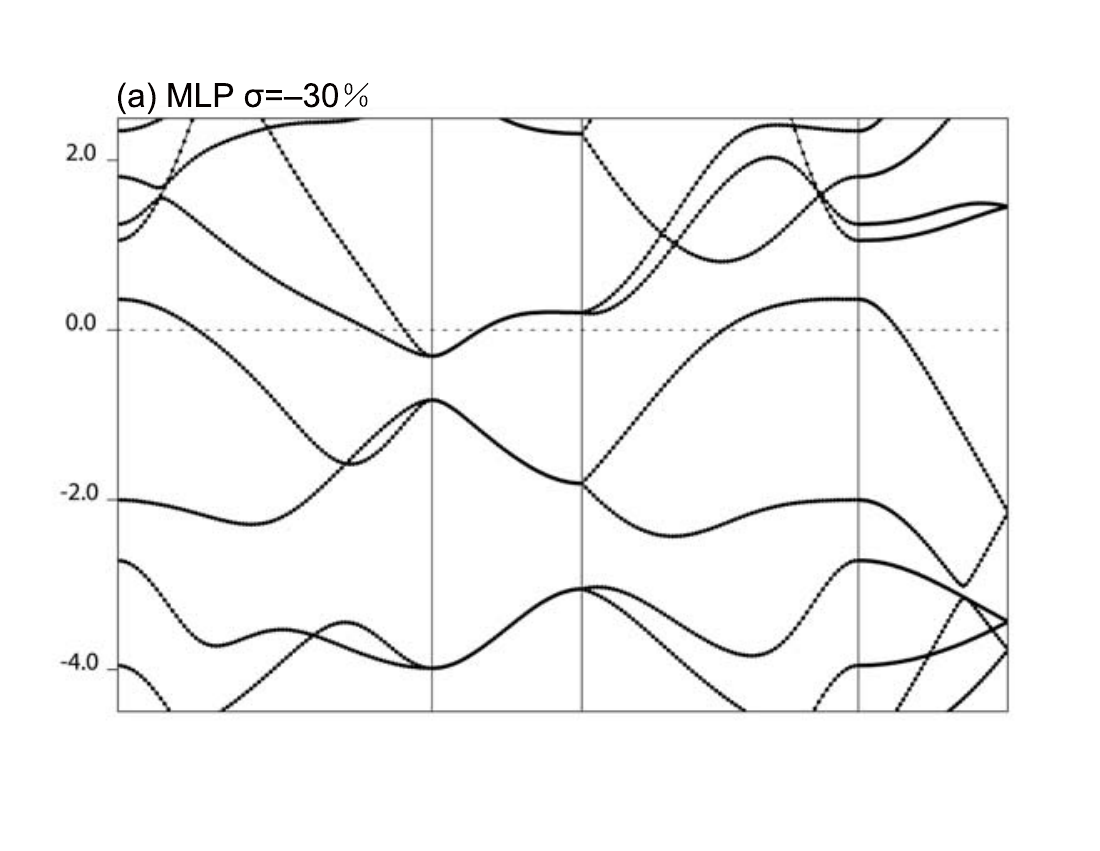}
} \hskip 1pt \subfigure {
\includegraphics[width=2.in]{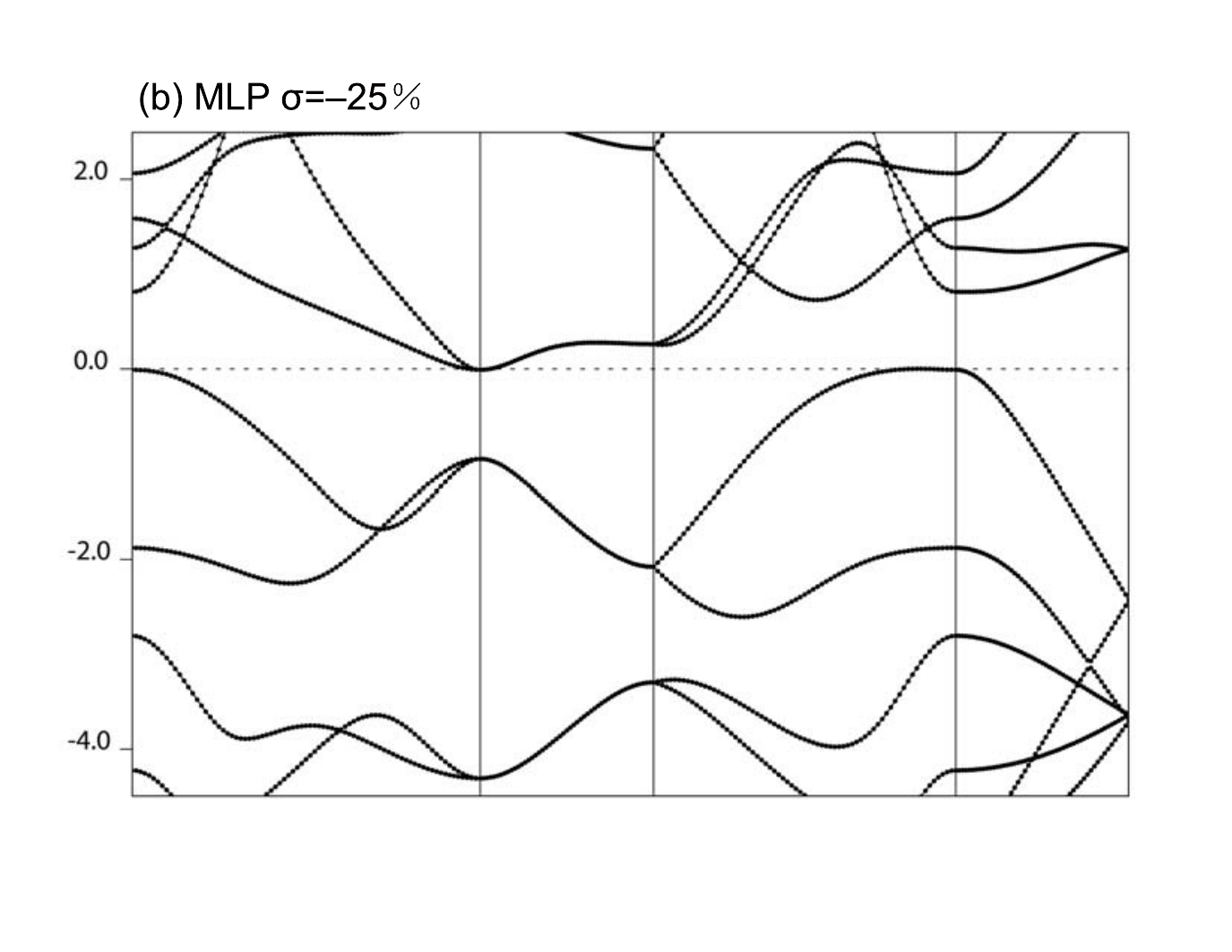}
}\hskip 1pt \subfigure {
\includegraphics[width=2.in]{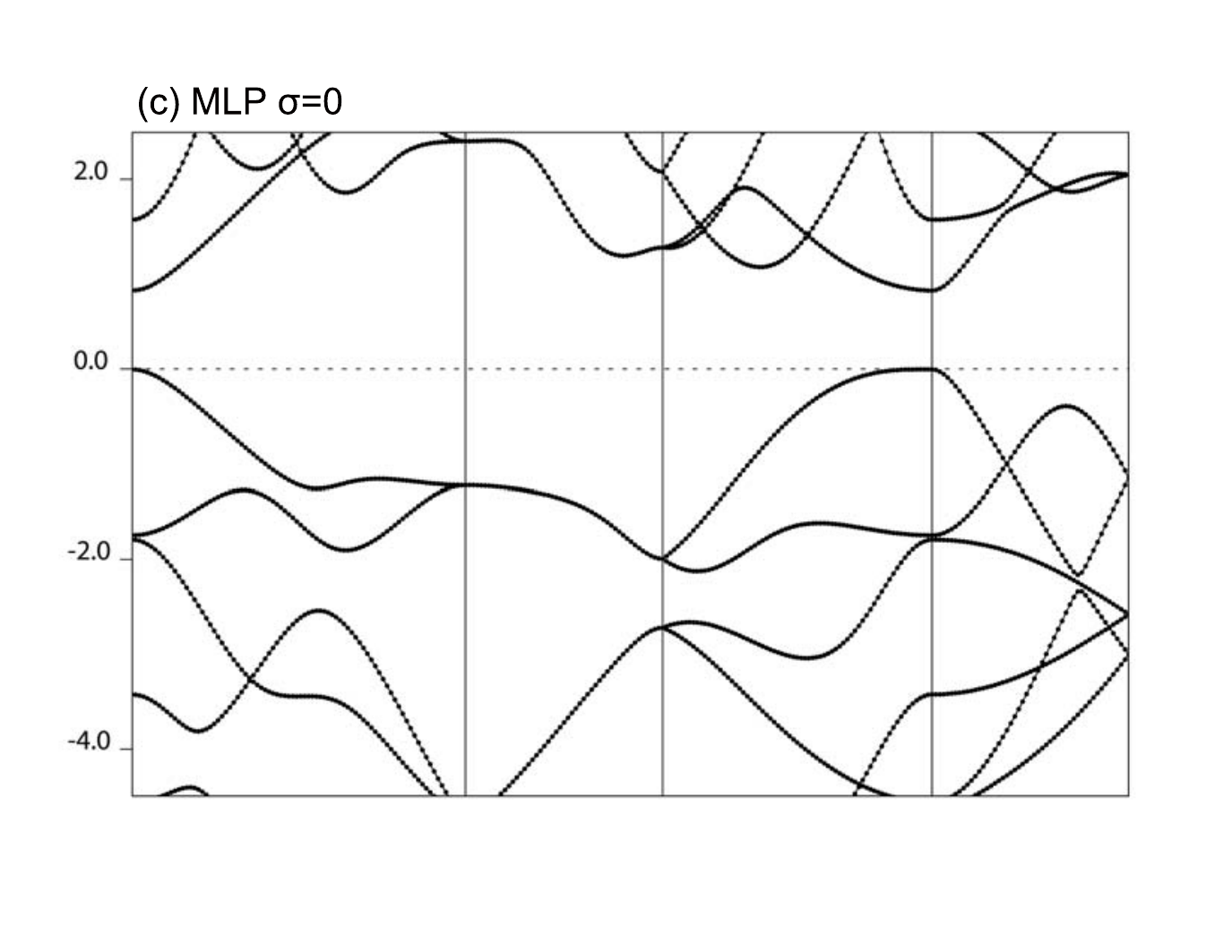}
}
\\
\subfigure {
\includegraphics[width=2.in]{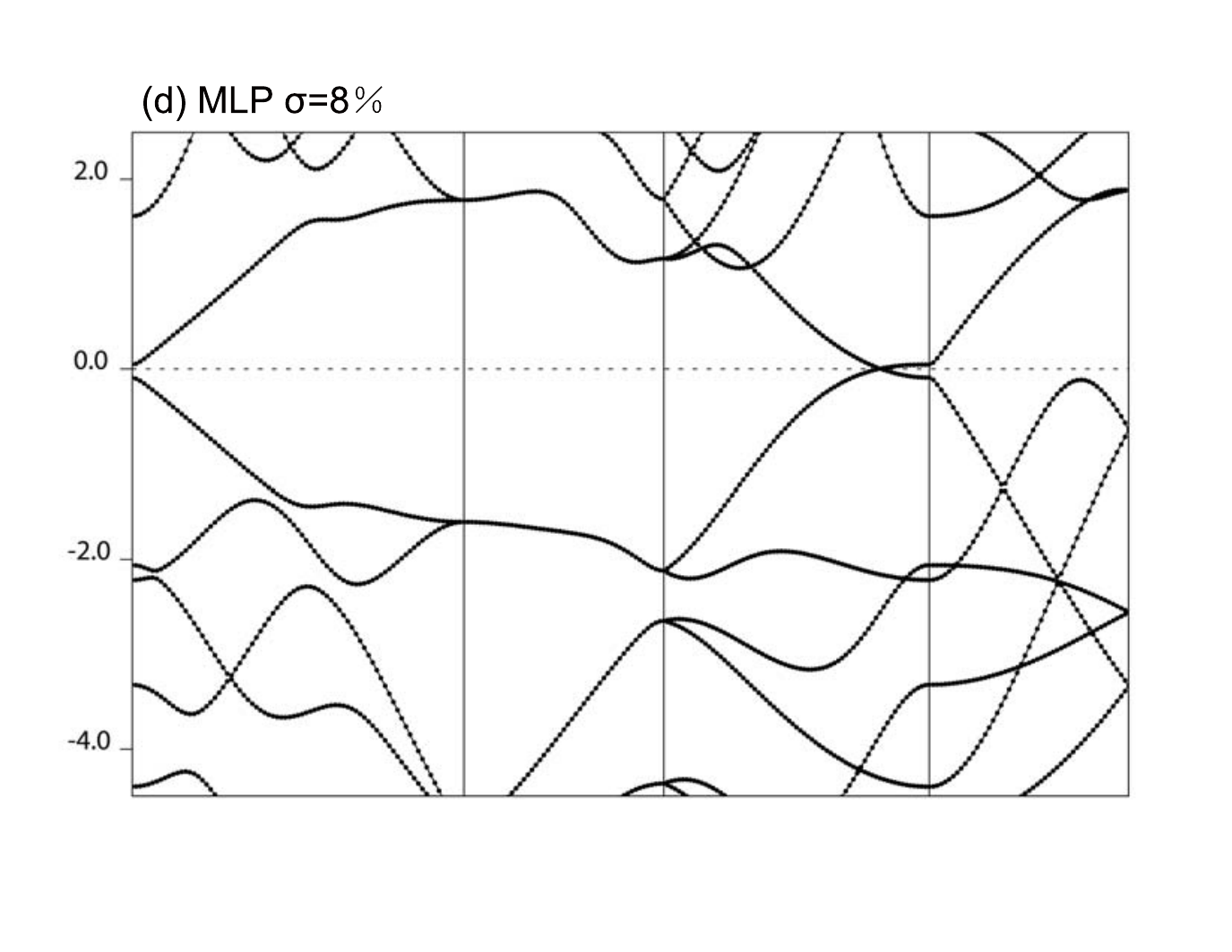}
} \hskip 1pt \subfigure {
\includegraphics[width=2.in]{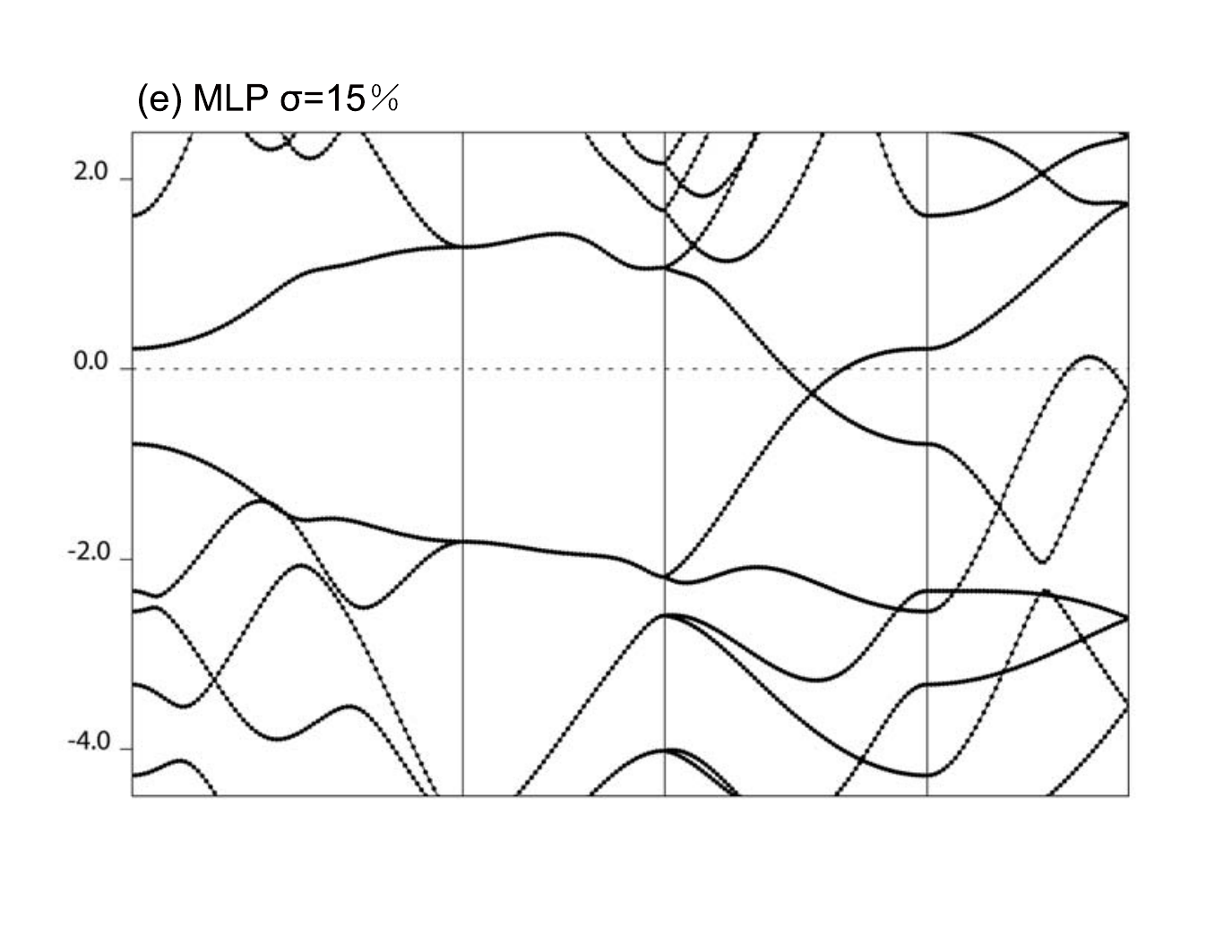}
}\hskip 1pt \subfigure {
\includegraphics[width=2.in]{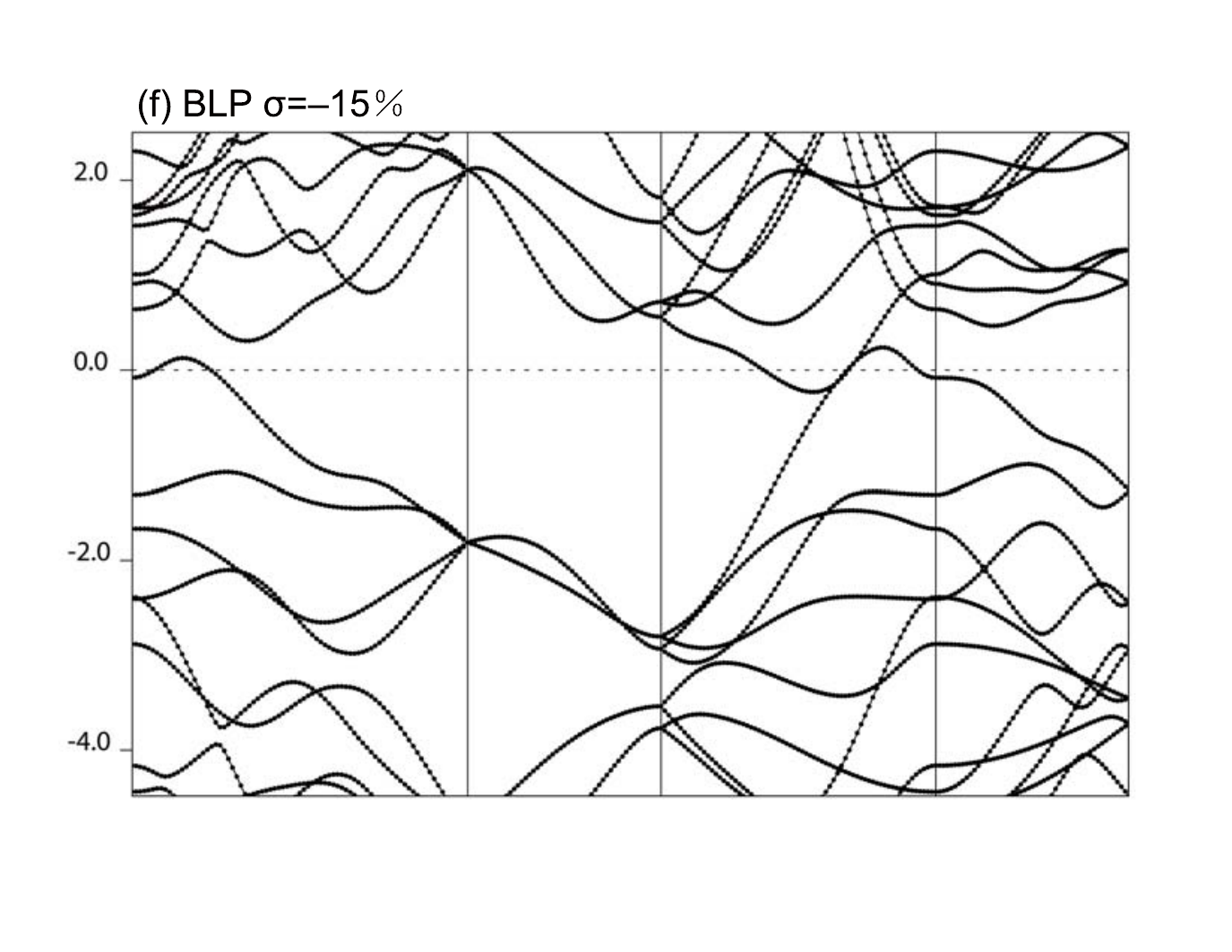}
}
\\
\subfigure {
\includegraphics[width=2.in]{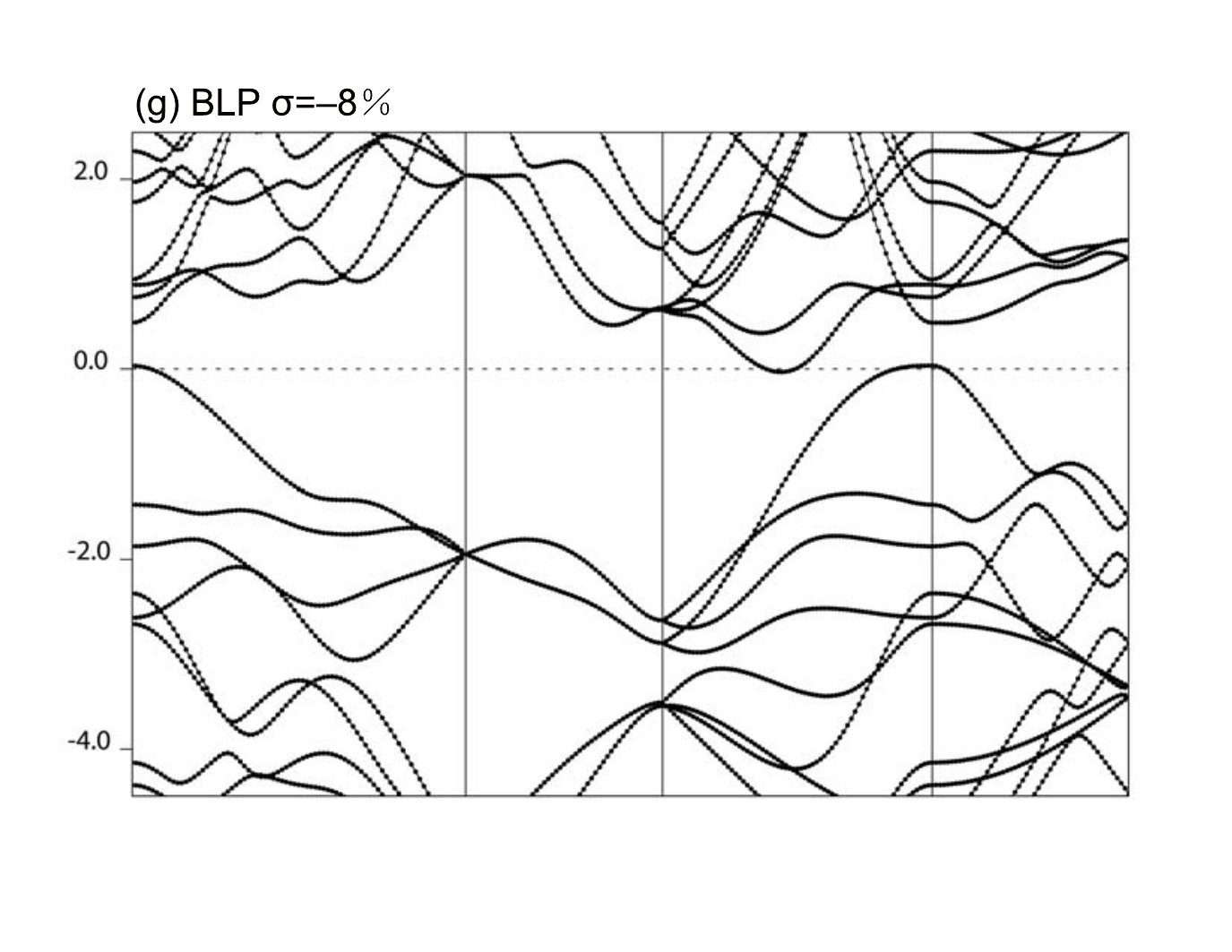}
} \hskip 1pt \subfigure {
\includegraphics[width=2.in]{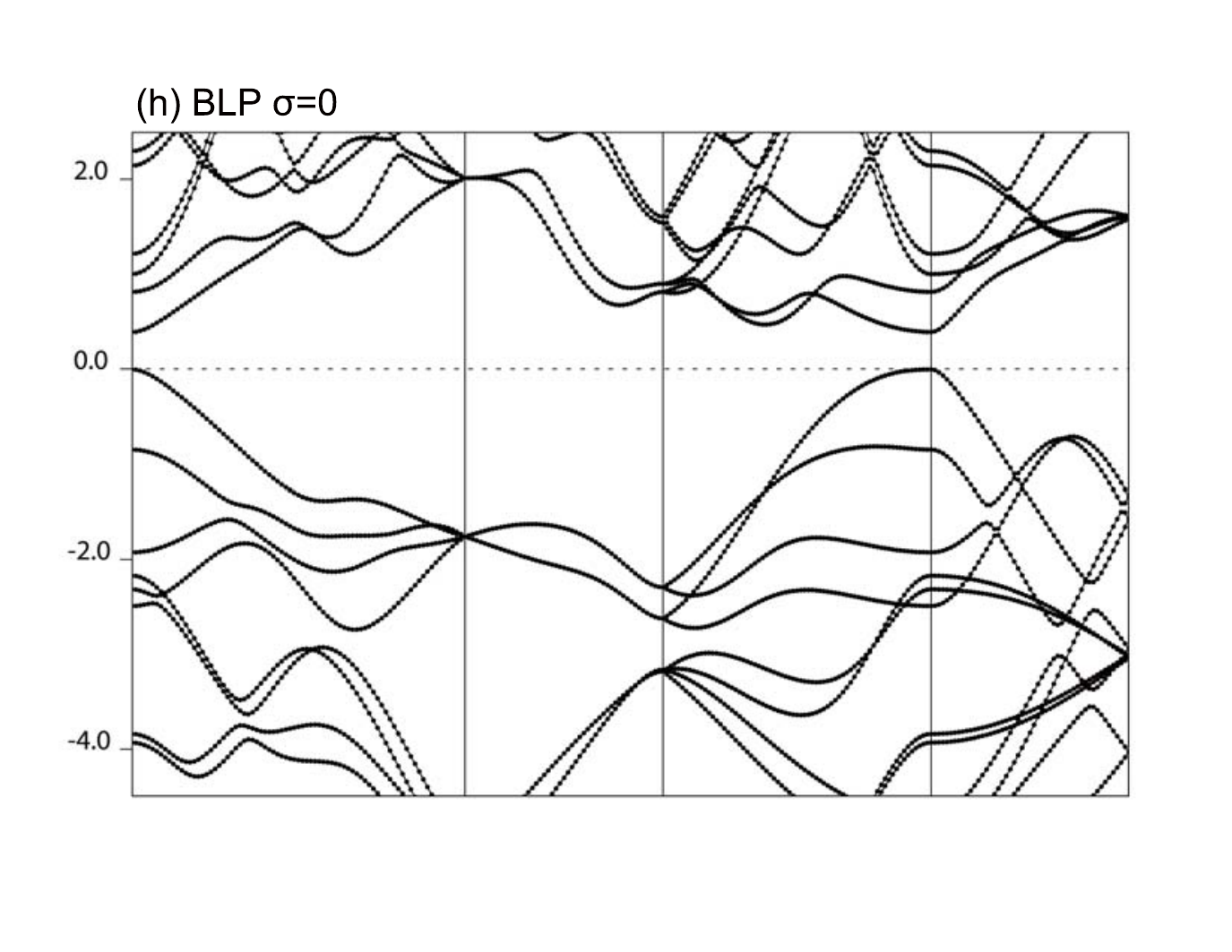}
}\hskip 1pt \subfigure {
\includegraphics[width=2.in]{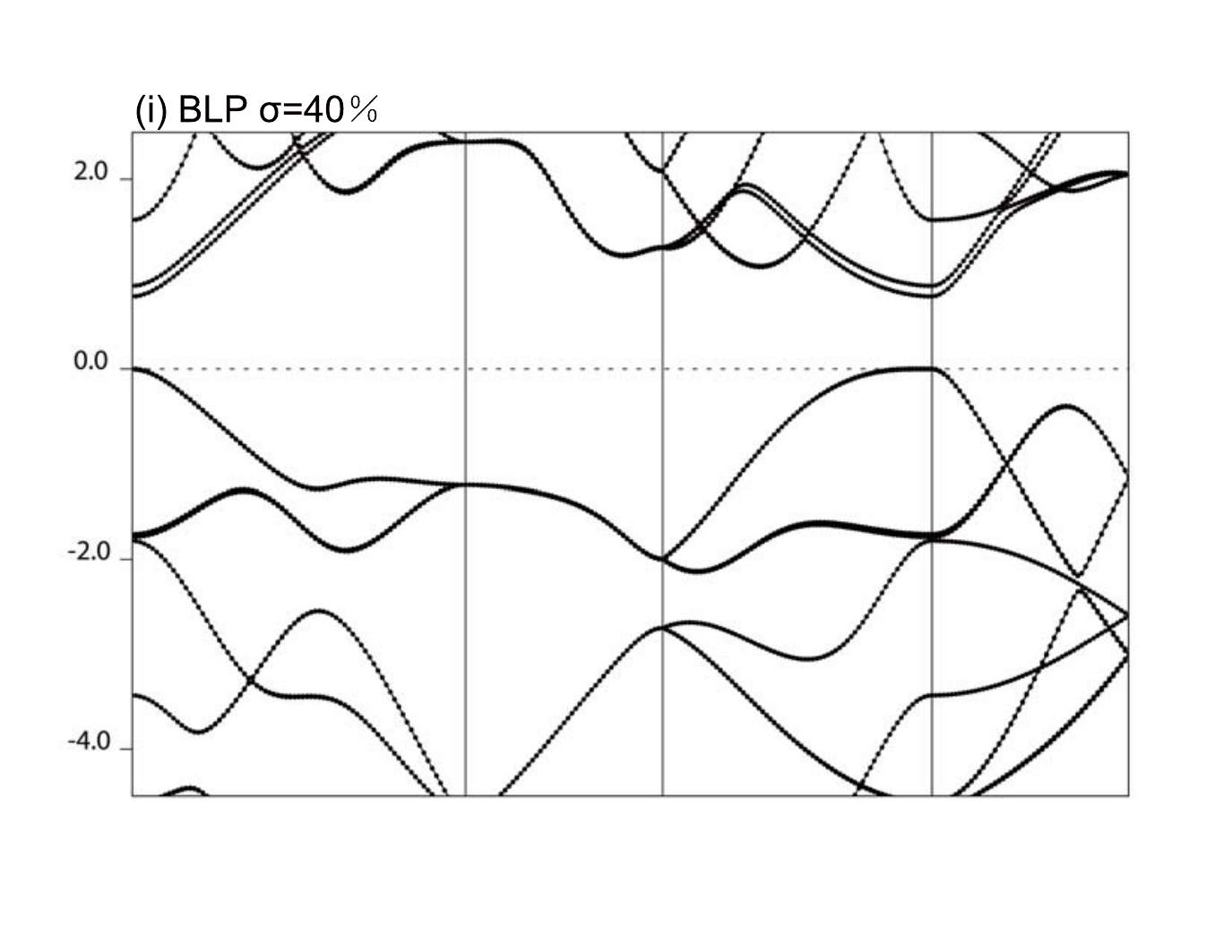}
}
\\
 \subfigure {
\includegraphics[width=2.2in]{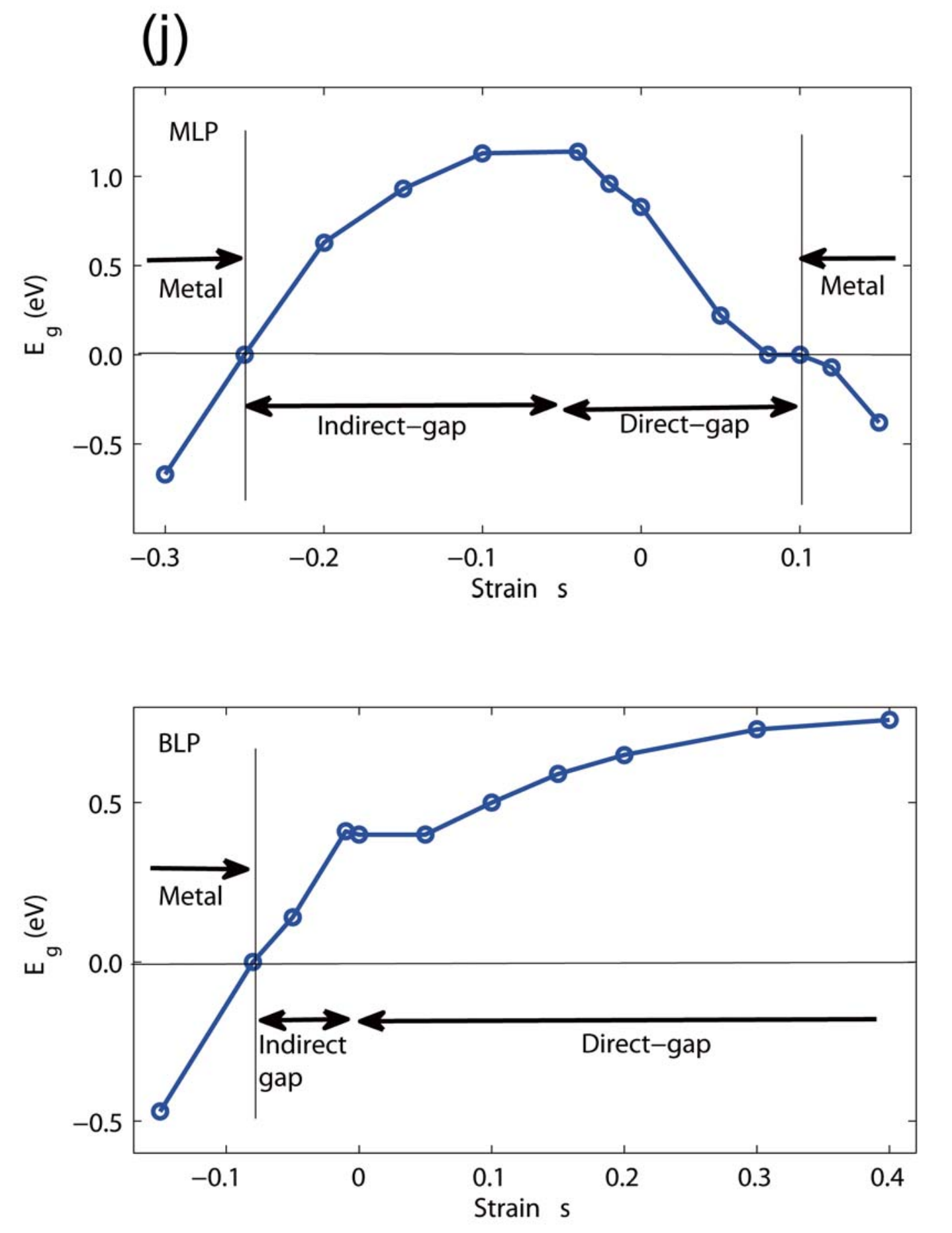}
}
\caption{The band structures of the strained  phosphorene, (a)-(e)
are for MLP; (f)-(i)are for BLP. (j) The evolution of the band gap
$E_{g}$ with the strain.}
\label{fig.2}
\end{figure}

\par
The vertical strain effect on the electronic structure of MLP can be
seen from (a) to (e) of Fig. 2. Both the compressive and tensile
strains applied for the MLP can result in the transition from
semiconductor to semimetal or metal. The band gap $E_{g}$ located at
$\Gamma$ point keeps direct and decreases with the tensile strain,
turning into zero when the tensile strain reaches $8\%$ (Fig. 2d),
and then converting into semimetal or metal (Fig. 2e) with
increasing the tensile strain continuously. When compressive strain
is applied, the band gap firstly keeps direct and increases
slightly, but turns into indirect when the compressive strain
reaches about $4\%$. Then indirect band gap decreases and turns into
zero at about $25\%$ compressive strain (Fig. 2b), and finally
converting into semimetal or metal when the compressive strain
continues to increase(Fig. 2a)

\par
For the BLP, when tensile strain is applied, the band gap $E_{g}$
keeps direct and increases until tends to the value in the free MLP.
The corresponding band structure (Fig. 2i) is also similar to that
of the free MLP (see Fig. 2c). Under small compressive strain
($1\%$), the band gap  soon turns into indirect and decreases with
increasing the compressive strain. It turns into zero when the
compressive strain reaches about $8\%$ (Fig. 2g), and finally
converts into metal (Fig. 2f).

\par
The evolution of the band gap $E_{g}$ with the strain for the MLP
and BLP is summarized in Fig. 2j. The change of  band gap is very
extensive. So we can find that the electronic structure of thin BP
is very sensitive to its thickness and the strain applied. This
properties help  to tailor materials in electronics. In order to
better understand this novel 2D materials and use them for potential
applications, next we turn to discuss the dynamical stability of the
strained phosphorene.

\subsection{ Lattice vibration under vertical strain}
\par
The symmetry of  phosphorene is described by the $D_{2h}$ point
group. For the MLP, the primitive cell contains four atoms, leading
to 12 vibrational modes, i.e. nine optical and three acoustic
phonons branches. The $\Gamma$ point modes can be decomposed as
\begin{equation}
\Gamma_{acoustic}=B_{1u}+B_{2u}+B_{3u}
\end{equation}
and
\begin{equation}
\Gamma_{optical}=B_{1u}+B_{3u}+A_{u}+B_{1g}+B_{3g}+2B_{2g}+2A_{g}
\end{equation}
Among optical modes, the $A_{u}$ mode is silent, $B_{1u}$ and
$B_{3u}$ are infrared active, while all others are  Raman active.
The structure of puckered layer remains the reflection symmetry in
the $y$ (zigzag) direction but  breaks the reflection symmetry in
the $z$ and $x$ directions. So the vibrations along the $y$
direction are strict, while those along the $z$ and $x$ directions
are mixed. $B_{3g}$ and $B_{1g}$  Raman modes  correspond to the
vibrations along $y$ direction and their vibrational patterns are
illustrated in Fig. 1c. For $B_{3g}$ mode, the vibration of atom in
the different layer is out-of-phase, but is in-phase in the same
layer, i.e. two layers beating against each other.  $B_{1g}$ mode
corresponds to the bond stretching modes from $d_{1}$ and $d_{2}$
covalent bonds. The other four Raman active modes have  similar
vibrational patterns along $x$ and $z$ directions, but each mode
having mixed $x$ and $z$ characters. The Raman frequencies of free
MLP are listed in Table 3.  Our calculated frequencies are slightly
smaller than those in Ref. [\onlinecite{16}]. For the free BLP, the
frequencies of corresponding Raman active modes are also listed in
Table 3. We find that the frequency shift between MLP and BLP is
small except for $A_{g}^{1}$ and $B_{2g}^{2}$ modes, which vibrate
mainly along the $z$ direction. The frequencies of this two modes in
BLP is smaller that those in MLP probably due to the attractive vdW
interaction existed in BLP.

\par
The phonon dispersion of  free MLP calculated by the density
functional  perturbation theory is presented in Fig. 3 (a), which
compares very well with previous calculations.\cite{10,16} Near the
$\Gamma$ point, two in-plane acoustic modes exhibit linear
dispersions, while the off-plane acoustic mode exhibits a parabolic
dispersion due to the 2D character of phosphorene. The vibrations of
the bond-breathing modes occupy  the high-frequency region, while
the vibrations  of layer-breathing modes and the acoustic modes
occupy the low-frequency region. Fig. 3 (b) shows the calculated
phonon dispersion of free BLP which has 24 phonon branches. The
optical branches have small splitting and are nearly double
degenerate due to the weak interlayer interaction. Near the $\Gamma$
point, the splitting has an appreciable magnitude. Three branches
from the opposite vibrations of two puckered layers (i.e.
layer-breathing modes) deviate from three acoustic branches. From
Fig. 3a and 3b, we can see that there is no imaginary frequency in
the full phonon spectra, indicating the dynamical stability for both
free-standing MLP and BLP. This conclusion is in good agreement with
the experiment. Raman spectroscopy and transmission electron
microscopy measurements show that  the exfoliated flakes of BP are
stable even in free-standing form.\cite{15}

\begin{figure}
\vskip 1in
\centering
\includegraphics[width=4.in]{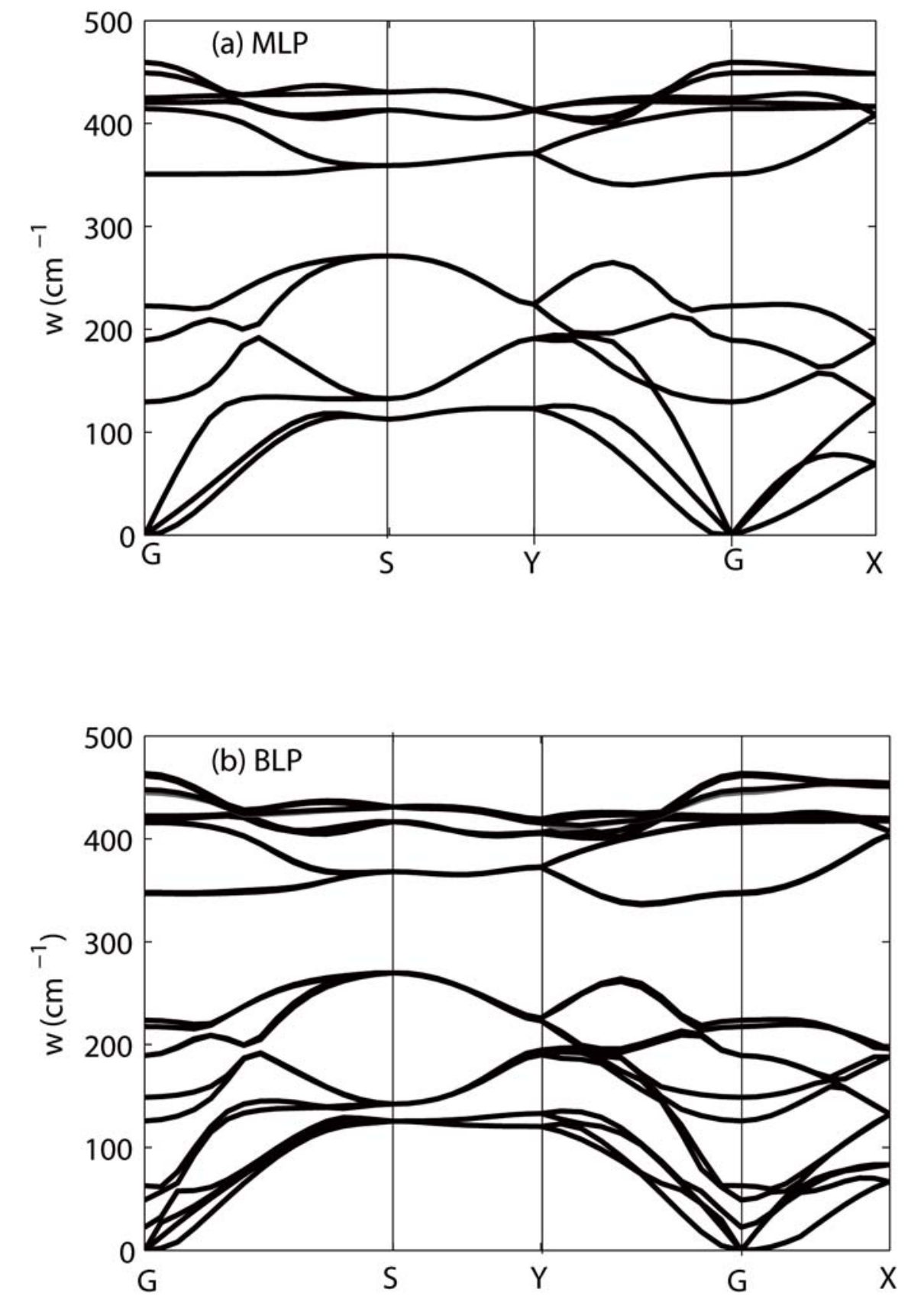}
\caption{The  phonon dispersion of free MLP (a) and BLP (b).}
\label{fig.3}
\end{figure}

\par
When vertical strain is increased to a certain value, we find
phosphorene becomes unstable. For MLP, there appears imaginary
frequency in the phonon dispersions when  tensile strain reaches
15$\%$ and compressive strain reaches 4$\%$. The appearance of
imaginary frequency of phonon modes indicates the dynamical
instability under such strain. The  dynamical stable range of MLP
for strain $\sigma$  is about ($-3\%,12\%$). Within this range, the
frequencies of Raman active modes under several strains are also
listed in Table 3. The response of each Raman mode to strain is not
the same. Four modes ( $B_{1g}, A_{g}^{1}, B_{2g}^{1},B_{2g}^{2}$)
exhibit monotonic change: their frequencies are red shifted under
tensile strain  and blue shifted under compressive strain. Among
them, $A_{g}^{1}, B_{2g}^{2}$ modes which are vibrated mainly along
the z direction have prominent shifts. The other two Raman modes
exhibit nonmonotonic change. $B_{3g}$ mode exhibits an abnormal red
shift under small compressive strain. This is possibly due to the
fact that the flattening of the puckered layer under compressive
strain is in favor of the opposite vibrations of two layers along
the zigzag chain. $A_{g}^{2}$ mode exhibits a nonmonotonic shift
under tensile strain. Its frequency decreases at small tensile
strain and then increases with increasing tensile strain. This
abnormal behavior can be understood from the following: $A_{g}^{2}$
mode corresponds to the vibration of $d_{1}$ bond-breathing mainly
along $x$ direction. The lattice constant $a$ is contracted when
vertical tensile strain is applied. When the effect of this
contraction of $a$ dominates over the effect of increase of
thickness of phosphorene, it will enhance the interatomic
interactions, resulting in the increase of frequency of $A_{g}^{2}$
mode.

\par
For BLP, we find its dynamical stable range for strain $\sigma$  is
wider than that of MLP. When compressive strain reaches 8$\%$,
imaginary frequency appears in the phonon dispersion along
$\Gamma$-X direction. Within the stable dynamical range for strain
$\sigma$, the  frequencies of Raman active modes are also listed in
Table 3. Under compressive strain, $B_{3g}$ mode exhibits an
abnormal red shift as in the case of MLP, while the frequencies of
other Raman modes are blue shifted as expected as usual. BLP is
still stable when  tensile strain reaches 40$\%$. As discussed
above, large tensile strain only results in large interlayer
distance with other geometric parameters in close to those  in  the
MLP. Then BLP may be regarded as two nearly independent MLPs, which
are stable in free-standing form as discussed above. Under the wide
range of tensile strain, the changes of frequencies of four modes (
$B_{3g},B_{1g}, A_{g}^{2}, B_{2g}^{1}$) are all very small. This
shows that the vertical tensile  strain has little effect on the
in-plane vibrational modes. With the increase of tensile strain,
$A_{g}^{1}, B_{2g}^{2}$ modes vibrated mainly along the $z$
direction are red shifted at first as expected, but they are then
blue shifted. This abnormal blue shift may be due to the decrease of
attractive vdW interaction at large interlayer distance.

\subsection{Superconductivity by adjusting the interlayer
distance} The above phonon calculations show that  BLP is dynamical
unstable when it achieves the transition from semiconductor to
metal. Here we propose that stable metal phase or even a BCS
superconductor can be achieved by adjusting the interlayer distance
of BLP. We fix the interlayer distance at different given values,
the unit cell and other atomic positions are then relaxed. Since the
results with increasing the interlayer distance are very similar to
those in the tensile strained BLP, next we only discuss the opposite
case.

\par
When  the interlayer distance  $0.6d_{0} <  d < d_{0}$ ($d_{0}$ is
the interlayer distance of free BLP),  our calculations show that
the lattice constant $a$ increases slightly, $d_{2}$ and
$\theta_{2}$ show only a weak variation, while the lattice constant
$b$,  $d_{1}$ and $\theta_{1}$ hardly change. This means that the
puckered  character of BLP remains and changes little at this range
of interlayer distance. When $d < 0.6d_{0}$, the structure of BLP
starts to change substantially. The calculations of band structure
show that the band gap $E_{g}$ goes to zero when $d = 0.85d_{0}$.
Keeping on decreasing  the interlayer distance will result in  the
transition from semiconductor to metal. The calculations of phonon
show that BLP is dynamical unstable when $d$ is $0.65d_{0}$ or less.
So the range of $d$ for the stable metal phase of BLP is about
$0.65d_{0} <  d < 0.85d_{0}$.

\par
The Eliashberg spectral function  $\alpha^{2}F(\omega)$ depends
directly on the EP matrix element,
$g_{{\bf{k}+\bf{q}}j^{'}{\bf{k}}j}^{\bf{q}\nu}$, which  can be
determined self-consistently using linear response theory. By
plotting $\alpha^{2}F(\omega)$  we can estimate the relative
strength of the EP coupling. The phonon density of states
$F(\omega)$ and $a^{2}F(\omega)$ are plotted in Fig. 4 for three
different interlayer distances, which are within the range of stable
metal phase of BLP. When $d = 0.8d_{0}$, $d = 0.75d_{0}$ and $d =
0.7d_{0}$, $F(\omega)$ is very similar, while $a^{2}F(\omega)$ is
different. With smaller interlayer distance, the Eliashberg spectral
function $a^{2}F(\omega)$ has a significantly enhanced peak at the
low-frequency side. The characteristic phonon modes which dominate
the  EP coupling are just from the opposite vibrations of two
puckered double layers. So we speculate that the interlayer vdW
attractive interaction may play an important role in enhancing the
EP coupling of BLP with smaller interlayer distance.

\begin{figure}
\vskip 1in
\centering
\includegraphics[width=4.in]{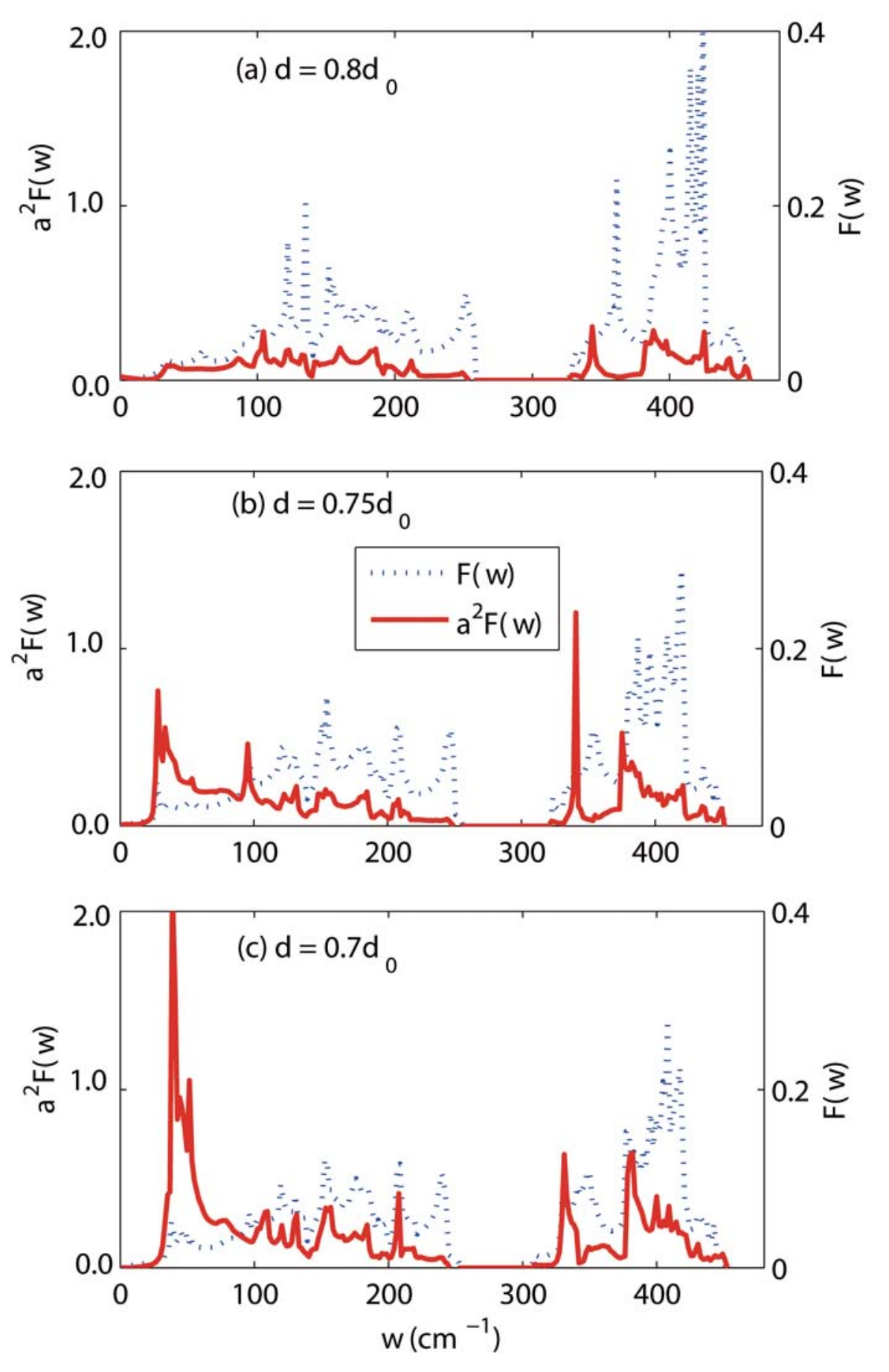}
\caption{The phonon density of states $F(\omega)$ and the
Eliashberg spectral function  $\alpha^{2}F(\omega)$ for BLP at three
different interlayer distances.}
\label{fig.4}
\end{figure}

\par
The superconducting temperature $T_{c}$ can be estimated from the
Allen-Dynes modified McMillan equation\cite{22}:
\begin{equation} T_{C}=
\frac{\omega_{ln}}{1.2}\exp
\left(-\frac{1.04(1+\lambda)}{\lambda-\mu^{*}(1+0.62\lambda)}\right),
\end{equation}
where $\omega_{ln}$ is the logarithmically averaged frequency,
$\mu^{*}$ is the Coulomb repulsion parameter. The calculated
electronic densities of states at Fermi level $N(E_{F})$, EP
coupling constant $\lambda$, $\omega_{ln}$ and the estimated $T_{C}$
when taking $\mu^{*}=0.1$ are summarized in Table 4. With the
decrease of interlayer distance, BLP  apparently becomes more
metallic  and results in the increase of $N(E_{F})$. Most
remarkably,  we see $\lambda$ increases dramatically.  The increase
of both $N(E_{F})$ and $\lambda$ is favorable to enhance $T_{C}$.
When $d = 0.7d_{0}$, $\lambda$ reaches as high as 1.45 and the
estimated $T_{C}$ is about 10K. The superconductivity of BP was
reported early under high pressure.\cite{5,6} Our results suggest
that BLP may become a good BCS superconductor by adjusting the
interlayer distance.

\section{Summary}
In conclusion, MLP and BLP  under vertical strain are studied by
density functional and  density-functional perturbation theory. The
results show that the electronic structure of thin BP is very
sensitive to its thickness and the strain applied.  The change is
extensive. It can increase direct band gap, decrease indirect band
gap or realize the transition from semiconductor to metal,  which
help to tailor materials in a variety of settings, from infrared
optoelectronics to high-mobility quantum transport. The shift of
zone-center Raman active modes for strained phosphorenes is analyzed
and explained combining the relaxation of geometric structure. Our
study clearly show that both MLP and BLP in free-standing form are
dynamical stable. However, material becomes dynamical unstable with
large vertical strain. We find that the dynamical stable range for
strain $\sigma$ in BLP is wider than that in MLP due to add the
interlayer weak vdW  coupling, this additional freedom. This
information is essential for future device fabrication and potential
applications of phosphorene. Furthermore we also find that BLP may
become a good BCS superconductor by adjusting the interlayer
distance.

\section{Acknowledgments}
The authors acknowledge the support of the Natural Science Foundation of Jiangsu Province in China under Grant No. BK20141441,the State Key Program for Basic Researches of China (2014CB921103 and 2010CB923404), the National "Climbing" Program of China (91021003), and the National Science Foundation of Jiangsu Province (BK2010012).

\newpage
\begin{center}{\bf Figure Caption}
\end{center}
Figure 1. The top (a) and  side views (b) of the atomic structure of
 phosphorene.  (c) Vibrational pattern of $B_{3g}$ and
$B_{1g}$ modes.\\
\\

Figure 2. The band structures of the strained  phosphorene, (a)-(e)
are for MLP; (f)-(i)are for BLP. (j) The evolution of the band gap
$E_{g}$ with the strain.\\
\\

Figure 3. The  phonon dispersion of free MLP (a) and BLP (b).\\
\\

Figure 4.  The phonon density of states $F(\omega)$ and the
Eliashberg spectral function  $\alpha^{2}F(\omega)$ for BLP at three
different interlayer distances.

\newpage

\begin{table}
\caption{Structural parameters of  bulk BP.}\label{tab.1}
\begin{center}
\begin{tabular}{cccccccc}
\hline\hline
&a(${\AA}$) &b(${\AA}$) &c(${\AA}$) &$d_{1}$(${\AA}$) &$d_{2}$(${\AA}$)&$\theta_{1}$ & $\theta_{2}$\\
 Exp. (Ref.\onlinecite{19})& 4.376 & 3.314& 10.478 & 2.224 & 2.244 & 96.34$^{\circ}$ & 102.09$^{\circ}$\\
 Theo.(Ref.\onlinecite{20})&4.422 & 3.348 & 10.587 & 2.238 & 2.261 & 96.85$^{\circ}$&
 102.31$^{\circ}$\\
This work (with vdW)&4.398 & 3.325 & 10.429 & 2.227 & 2.257 &
96.56$^{\circ}$& 102.20$^{\circ}$\\
This work (without vdW)&4.503 & 3.311 & 11.033 & 2.227 & 2.261 &
96.07$^{\circ}$&
 103.03$^{\circ}$\\
\hline\hline
\end{tabular}
\end{center}
\end{table}


\begin{table}
 \caption{ Structural parameters of  strained MLP and BLP. }
\label{tab.2}
\begin{center}
\begin{tabular}{c c c c c c c }
\hline\hline
$\sigma$&a(${\AA}$) &b(${\AA}$) &$d_{1}$(${\AA}$) &$d_{2}$(${\AA}$)&$\theta_{1}$ & $\theta_{2}$\\
&&&MLP&&&\\
 0.15 & 4.281 & 3.293& 2.198 & 2.545 & 97.04$^{\circ}$ & 100.27$^{\circ}$\\
 0.10 & 4.379 & 3.294& 2.210 & 2.451 & 96.38$^{\circ}$ & 101.24$^{\circ}$\\
 0.05 & 4.441 & 3.311& 2.224 & 2.355 & 96.22$^{\circ}$ & 102.04$^{\circ}$\\
 0.00 & 4.525 & 3.311& 2.222 & 2.255 & 96.35$^{\circ}$ & 103.35$^{\circ}$\\
 -0.05 & 4.671 & 3.302& 2.215 & 2.200 & 96.40$^{\circ}$ & 105.10$^{\circ}$\\
 -0.10 & 4.904 & 3.292& 2.201 & 2.159 & 96.81$^{\circ}$ & 107.74$^{\circ}$\\
 -0.25 & 5.729 & 3.287& 2.176 & 2.151 & 98.14$^{\circ}$ & 116.00$^{\circ}$\\
&&&BLP&&&\\

 0.40 & 4.527 & 3.308& 2.221 & 2.256 & 96.27$^{\circ}$ & 103.33$^{\circ}$\\
 0.15 & 4.523 & 3.308& 2.222 & 2.261 & 96.21$^{\circ}$ & 103.31$^{\circ}$\\
 0.10 & 4.513 & 3.310& 2.223 & 2.265 & 96.20$^{\circ}$ & 103.16$^{\circ}$\\
 0.05 & 4.486 & 3.311& 2.225 & 2.267 & 96.19$^{\circ}$ & 102.92$^{\circ}$\\
 0.00 & 4.485 & 3.314& 2.225 & 2.259 & 96.24$^{\circ}$ & 102.98$^{\circ}$\\
 -0.05 & 4.505 & 3.320& 2.226 & 2.233 & 96.46$^{\circ}$ & 103.33$^{\circ}$\\
 -0.10 & 4.607 & 3.324& 2.225 & 2.208 & 96.67$^{\circ}$ & 104.41$^{\circ}$\\
 -0.20 & 4.994 & 3.390& 2.220 & 2.199 & 99.55$^{\circ}$ & 107.91$^{\circ}$\\
\hline\hline
\end{tabular}
\end{center}
\end{table}

\newpage

\begin{table}
\caption{Frequencies ($cm^{-1}$)of Raman modes of strained
phosphorene. }\label{tab.3}
\begin{center}
\begin{tabular}{ccccccc}
\hline\hline

 $\sigma$ &$B_{3g}$& $B_{1g}$ &
 $B_{2g}^{1}$&$B_{2g}^{2}$&$A_{g}^{1}$&$A_{g}^{2}$ \\
&&&MLP\footnotemark[1] &&&\\
0.00&196&433&226&427&368&456\\
&&&MLP\footnotemark[2] &&&\\
0.12&170.1&417.4&152.5&276.6&147.8&455.6\\
0.10&175.2&415.8&162.8&297.1&152.0&444.0\\
0.08&180.9&416.2&175.9&320.1&170.1&433.5\\
0.05&187.2&418.1&196.0&357.8&266.9&431.4\\
0.00& 188.2&421.9&222.9&424.5&349.7&449.9\\
-0.02&187.8&424.3&226.8&437.5&354.5&459.2\\
-0.03&185.4&425.5&230.1&449.0&359.2&466.6\\
&&&BLP\footnotemark[2] &&&\\
0.40&188.9&422.3&221.9&423.4&349.5&449.5\\
0.15&189.0&421.7&221.4&420.3&345.4&448.2\\
0.10&189.1&421.2&221.1&418.0&344.9&448.5\\
0.05&188.9&421.2&220.0&414.7&341.5&446.2\\
0.00&188.1&422.6&221.2&419.9&346.7&450.3\\
-0.03&186.2&424.2&226.3&431.0&353.6&452.6\\
-0.05&185.8&426.0&228.1&436.7&357.8&463.0
\\
\hline\hline
\end{tabular}
\footnotetext[1]{Ref.~\onlinecite{16}.} \footnotetext[2]{This work}
\end{center}
\end{table}


\begin{table}
\caption{The calculated  electronic densities of states at Fermi
level $N(E_{F})$, EP coupling constant $\lambda$,  the
logarithmically averaged frequency $\omega_{ln}$ and the estimated
$T_{C}$ for BLP at three different interlayer distances.}
\label{tab.4}
\begin{center}
\begin{tabular}{ccccc}
\hline\hline

d/$d_{0}$ &$N(E_{F})(States/eV)$& $\lambda$ &$\omega_{ln}(K)$&$T_{C} (K)$\\
0.80& 0.68&0.32&183.2&0.2\\
0.75& 0.86&0.85&107.9&5.7\\
0.70 & 1.08&1.45&96.1&10.6\\

\hline\hline
\end{tabular}
\end{center}
\end{table}

\end {document}